\documentstyle[preprint,eqsecnum,aps]{revtex}
\begin{document}
\preprint{\it\vbox{\hbox{DOE/ER/40762-163}\hbox{U.~of MD PP\#99-038}}}
\title{Discretization Errors and Rotational Symmetry:\\
The Laplacian Operator on Non-Hypercubical Lattices}
\author{Chi--Keung Chow}
\address{Department of Physics, University of Maryland, College Park, 
20742-4111.}
\date{\today}
\maketitle
\begin{abstract} 
Discretizations of the Laplacian operator on non-hypercubical lattices are 
discussed in a systematic approach.  
It is shown that order $a^2$ errors always exist for discretizations  
involving only nearest neighbors.  
Among all lattices with the same density of lattice sites, the hypercubical 
lattices always have errors smaller than other lattices with the same 
number of spacetime dimensions.  
On the other hand, the four dimensional checkerboard lattice (also known 
as the Celmaster lattice) is the only lattice which is isotropic at order 
$a^2$. 
Explicit forms of the discretized Laplacian operators on root lattices are 
presented, and different ways of eliminating order $a^2$ errors are discussed.
\end{abstract}
\widetext

\section{Introduction}

Lattice simulation is an invaluable tool in our studies of nonperturbative 
aspects of quantum field theories.  
By the discretization of the spacetime continuum, the original system with 
infinitely many degrees of freedom is approximated by one with only a large 
but finite number of degrees of freedom, which can be studied numerically.  
In practice, the lattice has almost always been chosen as hypercubical since 
the original paper by Wilson \cite{Wil}, although some ``uncanonical'' choices 
has also been discussed.  
Four-dimensional random lattices were discussed by Christ {\it et.~al.}~in 
Ref.~\cite{CFL1,CFL2,CFL3}.  
The body-centered hypercubical lattices in four dimensions had been first 
studied by Celmaster \cite{Cel1,Cel2,Cel3} and subsequently in 
\cite{DMP,Neu1,Neu2,Neu3}.  
Almost at the same time four dimensional simplicial lattices were studied 
in \cite{DM1,DM2,DM3,DM4,DM5}.  
(An interesting comparison between the last two uncanonical lattices can be 
found in Ref.~\cite{CM}.)

Let's recall the definition of a lattice.  
A set of vectors $\{B_i\}$ ($1\le i \le n$) in $m$-dimensional 
space ${\bf R}^m$ is {\it linearly independent} if 
\begin{equation}
\sum_{i=1}^n x_i B_i = 0 \qquad {\rm iff} \qquad x_i = 0  \quad \forall i.
\end{equation}
With such a set of linearly independent basis vectors $\{B_i\}$ one can 
define a lattice as below.  

\bigskip

{\sl An $n$-dimensional lattice $\Lambda_n$ is a subset of 
$\,{\bf R}^m$ $(n\le m)$ such that 
\begin{equation}
\Lambda_n = \{X=\sum_{i=1}^n z_i B_i: z_i\in {\bf Z}\}
\end{equation}}

\noindent The lattice always contains the origin and is symmetric under 
reflection $X \to -X$ in ${\bf R}^m$.  
Notice that we have not required that $n=m$ as some lattices (like the 
triangular lattice in two dimensions) can be more conveniently represented 
when embedded in some higher dimensional spaces.  
Also note that, under this definition, random lattices are {\it not\/} 
lattices.  
Still, such a definition encompasses a huge set of lattices, including 
extremely skewed ones whose unitcells are long narrow parallelograms or 
parallelohedra.
However, such highly asymmetric lattices are rarely used for simulations.  
Instead, physicists always use lattices which are ``highly symmetric'' 
(we will give a mathematically precise meaning to this phrase later).  
In particular, there is  a class of lattices called the ``root lattices'', 
which contains most of the ``highly symmetric'' lattices, at least in lower 
dimensions $n\le4$.  
While there is no real justification in neglecting the lattices which do 
not belong to this class, root lattices are so well studied that they provide 
a natural starting point for our investigation.  
As we will see, there are exactly three root lattices in four dimensions, 
namely the hypercubical, the body-centered hypercubical and the simplicial 
lattices\footnote{This fact has been observed in Ref.~\cite{DM1}.}, 
{\it i.e.}, exactly the cases which have been studied.  

It is important to compare the relative merits of these different lattices 
in simulations.  
Several papers mentioned above attempted to answer this question by 
simulating the same physical system on different lattices and comparing the 
results.  
These comparisons are very directly relevant to the practical applications 
of lattice simulations, but theoretically they are ``unclean'' in the 
sense that they are {\it a priori\/} sensitive to the physical problems in 
question, the actual algorithms used, and other non-geometrical factors.  
This paper attempts to study the same problem in a complementary way.  
We will obtain a general prescription on how to construct the Laplacian 
operator, the simplest possible differential operator, on any lattice. 
While the Laplacian operator is not directly relevant for lattice QCD, 
it is used in lattice simulations of strongly coupled scalar $\phi^4$ 
theories, and it is logical to study the Laplacian case thoroughly before 
tackling the more complicated cases of the Dirac and the plaquette 
operators.  
By studying the discretization errors of the Laplacian operators 
mathematically, one can parameterize these errors in ways which depend only 
on the geometrical properties of the lattices.  
The questions we will be addressing include: 

$\bullet$ Are there any lattices on which the discretizations of the 
Laplacian operator on the nearest neighbors are free of ${\cal O}(a^2)$ 
errors?  

$\bullet$ What are the merits and demerits of our canonical choice --- 
the simple hypercubical lattice?  

$\bullet$ What is the role of symmetry? 

$\bullet$ In general, how do we eliminate the ${\cal O}(a^2)$ errors?  
(Answer: by involving sites which are not nearest neighbors.)  
Is it always possible?  

$\bullet$ In a given dimension, does the lattice which has the smallest 
${\cal O}(a^2)$ errors still work best after the ${\cal O}(a^2)$ errors 
are cancelled?  

In this paper, we will try to provide answers to all these questions.  
The approach is ``clean'' in the sense that all the effects of algorithms 
or choices of physics problems are disentangled.  
On the other hand, it must be admitted that its relevance to practical 
lattice simulations is not immediate.   
Roughly speaking, this paper is more about lattice field theory than 
about lattice simulations, and the theorems obtained should be regarded as 
general ``rules of thumb'' instead of infallible laws in practice.  

This paper is organized as follows: 
In Section II we present a prescription (one shell discretization) 
for approximating the Laplacian on a lattice.  
Root lattices are defined in Section III, and the explicit forms of the 
discretized Laplacian operators on root lattices are presented in Section IV.  
Order $a^2$ errors are studied in Sections V and VI, while the issue of 
anisotropy is discussed in Section VII.  
In Section VIII a general prescription (two shell discretization) for 
eliminating the order $a^2$ errors is presented, and the order $a^4$ errors 
in this scheme are estimated in Section IX, followed by the conclusion in 
Section X.  

\section{Laplacian on a Lattice}

On hypercubical lattices, the discretization of the Laplacian operator is 
usually constructed through the trapezoidal rule applied on each lattice 
direction.  
For example, the second derivative in the $x$ direction of an arbitrary 
function $\phi$ at the origin is 
\begin{equation}
a^2 \partial_x^2 \phi(0) = \phi(a\hat x) - 2\phi(0) + \phi(-a\hat x) 
+ \dots, 
\end{equation}
and by summing up over all lattice directions, 
\begin{eqnarray}
a^2 \partial^2 \phi(0) &=& \sum_{\hat\mu} \Big(\phi(a\hat\mu) - 2\phi(0) 
+ \phi(-a\hat\mu)\Big) + \dots\nonumber\\
&=& \bigg\{ \sum_{\hat\mu} \Big(\exp(a\partial_\mu) -2 + \exp(-a\partial_\mu)
\Big)\bigg\} \phi(0) + \dots,  
\label{Z}
\end{eqnarray}
where the $\hat\mu$'s are the unit vectors along the lattice directions.  

The construction above depends so heavily on the existence of a set of 
mutually orthogonal lattice vectors $\{\hat\mu\}$ that it cannot be naturally 
generalized to non-hypercubical lattices.  
(Try the two dimensional triangular lattice, for instance.)  
To obtain a construction which is equally applicable to hypercubical and 
non-hypercubical lattices, we will make use of the coordinate-independent 
definition of the Laplacian operator.  
Recall that when $\phi$ satisfies the $n$-dimensional Laplace equation 
$\partial^2\phi=0$, the value of $\phi$ at any point is the average of its 
value on an $(n-1)$-dimensional hypersphere of arbitrary radius centered at 
the point in question.  
\begin{equation}
\phi(0) = \int_{S^{n-1}} \; d\mu_{n-1} \; \phi(X), 
\label{S}
\end{equation}
where $d\mu_{n-1}$ is the unit measure on $S^{n-1}$.  
\begin{equation}
\int_{S^{n-1}} \; d\mu_{n-1} \equiv \int_{S^{n-1}} \; d\Omega_{n-1} 
{\Gamma(n/2) \over 2\pi^{n/2}} = 1.  
\end{equation}
So one can take the Laplacian $\partial^2\phi$ as an indicator of how much 
the equality Eq.~(\ref{S}) is violated.  
Just from dimensional analysis one knows that 
\begin{equation}
\partial^2\phi(0) = N \lim_{r\to0} \bigg( \int_{S^{n-1}} \; d\mu_{n-1} \; 
\phi(X) - \phi(0) \bigg)\bigg/r^2, 
\label{DefL}
\end{equation}
where $r$ is the radius of the hypersphere $S^{n-1}$, and $N$ is a 
proportionality constant one can easily determine to be $2n$.\footnote
{The value of $N$ can be determined by considering the following 
electrostatic system satisfying the Poisson equation $\partial^2\phi=\rho$ 
in $n$ dimensions.
Consider a ball of constant charge density $\rho$ and radius $r$.  
One can easily see that the potential difference between the surface and 
the center of the ball is $\rho r^2/2n$.  
Substitute this back into Eq.~(\ref{DefL}) and $N$ is found to be $2n$.}  

On a lattice, however, $\phi$ is only defined on discrete lattice sites, 
and one can at best approximate the integral with the sum of $\phi$ over 
discrete points.  
It is useful to introduce some formalism here and define a shell ${\cal S}_r$ 
of lattice sites with radius $r>0$ as the set of all sites with $\|X\| = r$.  
Obviously these ${\cal S}_r$'s are non-empty only for discrete values of 
the $r$'s, and the smallest of these values, $\rho$, is the separation between 
each lattice site and its nearest neighbors and is usually called the packing 
diameter of the lattice.  
We will call ${\cal S}_\rho$ the first shell, and its cardinality, {\it 
i.e.}, the number of nearest neighbors, denoted by  $\tau$, is called the 
kissing number\footnote
{Also known to chemists as the contact number or the coordination number, 
and to historians as the Newton number, after Isaac Newton who studied the 
problem in three dimensions.}.  
Now we approximate the integral of $\phi$ over $S^{n-1}$ by its average 
$\Sigma$ over the first shell
\begin{equation}
\Sigma \phi(0) = {\textstyle{1\over\tau}} \sum_{X_k\in{\cal S}_\rho} \; 
\phi(X_k), 
\end{equation}
where $\Sigma$ is defined as an operator over $\phi$, which can also be 
expressed in terms of the partial derivatives.  
\begin{equation}
\Sigma \phi(0) = \bigg\{ {\textstyle{1\over\tau}} \sum_{X_k\in{\cal S}_\rho} 
\; \exp(X_k \cdot \partial) \bigg\} \; \phi(0).   
\end{equation}

The corresponding approximation of the Laplacian operator is 
\begin{equation}
D^2\phi(0) = 2n \bigg( {\textstyle{1\over\tau}} \sum_{X_k\in{\cal S}_\rho} \; 
\phi(X_k)- \phi(0) \bigg)\bigg/\rho^2 = 2n (\Sigma - 1) \phi(0)/\rho^2, 
\label{DefD}
\end{equation}
The nonlocal operator $D^2$ defined above, called the {\it one shell 
discretization\/} of the Laplacian operator, is the main object of this 
study.  
How good is $D^2$ as an approximation of $\partial^2$?  
Note that 
\begin{eqnarray}
\rho^2 D^2\phi(0)/2n &=& {\textstyle{1\over\tau}} \sum_{X_k\in{\cal S}_\rho} 
\; \phi(X_k) - \phi(0) = {\textstyle{1\over\tau}} \sum_{X_k\in{\cal S}_\rho} 
\; \Big(\phi(X_k) - \phi(0)\Big)\nonumber\\&=&\bigg\{{\textstyle{1\over\tau}} 
\sum_{X_k\in{\cal S}_\rho} \; \Big(\exp(X_k \cdot \partial) - 1 \Big) \bigg\}
\; \phi(0).   
\end{eqnarray}
So the expression inside the braces is a derivative representation of the 
nonlocal operator $\rho^2 D^2/2n$.  
By expanding the exponential, which is equivalent to a Taylor expansion in 
$\rho$, one can express $\rho^2 D^2/2n$ in terms of powers of the derivative 
operator.  
Now 
\begin{equation}
\exp(X_k \cdot \partial) = 1 + X_k \cdot \partial + \textstyle{1\over2} 
(X_k \cdot \partial)^2 + \textstyle{1\over6} (X_k \cdot \partial)^3 
+ {\cal O}(\rho^4) .  
\end{equation}
The order $\rho^0$ term is just unity, which is cancelled by the ``$-1$'' 
coming from the $\phi(0)$ term in Eq.~(\ref{DefD}).  
The order $\rho^1$ term is proportional to 
\begin{equation}
\Big( \sum_{X_k\in{\cal S}_\rho} X_k \Big) \cdot \partial = 0 , 
\end{equation}
due to the reflection symmetry of the lattice.  
Similarly, all odd powers of $\rho$ vanish due to reflection symmetry.  
On the other hand, the $\rho^2$ term is ${1\over2}\sum_k(X_k \cdot \partial)^2$
which is in general {\it not\/} a Laplacian unless the set of nearest 
neighbors ${\cal S}_\rho$ has vanishing quadrapole moment.  
So we have the theorem: 

\bigskip

{\sl The one shell discretization $D^2$ of the Laplacian operator over the 
first shell ${\cal S}_\rho$ is equal to the Laplacian operator $\partial^2$ 
itself in leading order if and only if the quadrapole moment of 
${\cal S}_\rho$ vanishes.} 

\bigskip

It must be emphasized that the theorem above applies only for one shell 
discretizations as defined in Eq.~(\ref{DefD}).  
In general one may give different weights to different lattice sites in 
${\cal S}_\rho$ in the sum in Eq.~(\ref{DefD}) or even extend the sum to 
sites which are not nearest neighbors, and the theorem does not apply to 
such constructions.  
However, such alternative discretization schemes contain many free 
arbitrary parameters (as in the weights) and they are clearly beyond the 
scope of this paper.  

\section{Root Lattices} 

We have seen that the one shell discretization method provides a good 
approximation of the Laplacian operator as long as the nearest shell of 
neighbors has vanishing quadrapole moment.  
Unfortunately, the author is not aware of any classification of the 
lattices satisfying this condition.  
However, it is well known that root lattices have been completely 
classified, and all of them have vanishing quadrapole moment.  
Moreover, both the body-centered hypercubical and the simplicial lattices 
fall into this class, and so does the usual hypercubical lattice.  
So it is useful to study these root lattices closely.   

\bigskip

{\sl A root lattice is one whose basis vectors can all be chosen to be nearest 
neighbors and the angles between any two of them are either $\pi/2$ or 
$2\pi/3$.   
In other words, when the basis vectors are scaled to have unit length, the 
inner products of any two of them are either $0$ or $-{1\over2}$.  }

\bigskip

\noindent This definition depends solely on the relative directions of 
the nearest neighbors (as opposed to definitions in textbooks on Lie groups, 
which may involve next-to-nearest neighbors) is more useful for the study 
of one shell discretizations, which also just depend on the first shell.  

The classification theorem for root lattices is as follows: 

\bigskip

{\sl An $n$-dimensional root lattice is either isometric to one of the 
Cartan--Dynkin root lattices $Z_n$, $A_n$, $D_n$ and $E_n$ for some 
value of $a$, or is an outer product of several of them with the same $a$.  }

\bigskip

\noindent The Cartan--Dynkin lattices $Z_n$, $A_n$ and $D_n$ are defined 
as below for any $n \in {\bf Z}$:  
\begin{eqnarray}
Z_n \equiv \{X=(x_1,\dots,x_n)& &\in a{\bf Z}^n \}, \nonumber\\
A_n \equiv \{X=(x_1,\dots,x_n,&x_{n+1})\;&\in a{\bf Z}^{n+1}; \;
\sum_k x_k = 0\},\label{ZAD} \\
D_n \equiv \{X=(x_1,\dots,x_n)& &\in a{\bf Z}^n; \;\sum_k x_k \in 
2a{\bf Z}\},
\nonumber
\end{eqnarray}
These representations have the nice features that the coordinates of the 
lattice sites are always integer multiples of the lattice spacing $a$, and 
each lattice is symmetric under permutations of the coordinates.  
It is obvious that $Z_n$ are the square, cubic and hypercubical lattices for 
$n=2$, 3 and 4 respectively.  
The simplicial lattices $A_n$ are for $n=2$ the hexagonal lattices, for 
$n=3$ the face-centered cubic lattices, and for $n=4$ the four dimensional 
simplicial lattices studied in Ref.~\cite{DM1,DM2,DM3,DM4,DM5}.  
The checkerboard lattices $D_n$ are also square lattices for $n=2$ 
$(D_2 \simeq Z_2)$, face-centered cubic lattices for $n=3$ $(D_3 
\simeq A_3)$ and Celmaster lattices \cite{Cel1,Cel2,Cel3} for $n=4$.  
Lastly, the $E_n$ lattices are defined only for $n=6$, 7 or 8.  
Since we are mainly interested in lattices in low dimensions, we will 
not discuss $E_n$ and refer the interested reader to 
Ref.~\cite{Con} instead.  

Several questions immediately come to mind.  
Firstly, are the $A_n$ and $D_n$ really the same Cartan--Dynkin root 
lattices one uses for the representations of Lie groups\footnote{
Readers who are not familiar with the Cartan--Dynkin classification 
scheme for Lie groups can consult Slansky \cite{Sla} and Georgi \cite{Geo}.}?  
What is $Z_n$ in that language, and where are the $B_n$, $C_n$, and the 
exceptional groups?  
The first question can be easily answered by evaluating the angles between 
the basis vectors of $A_{n-1}$: 
\begin{equation}
(x_1, \dots, x_n) = (1,-1,0,\dots,0), (0,1,-1,0,\dots,0), 
\dots, (0,\dots,0,1,-1),
\end{equation}
where each vector is rotated by an angle $2\pi/3$ from its predecessor as 
described by the $A_n$ Dynkin diagrams.  
The $D_n$ lattice has one more basis vector than $A_{n-1}$, namely $(-1,-1,0,
\dots,0)$, which is orthogonal to the first basis vector of $A_{n-1}$ but 
makes an angle of $2\pi/3$ with the second, exactly as decreed by the $D_n$ 
Dynkin diagrams.  
Lastly, the $Z_n$ lattice also has one more basis vector than $A_{n-1}$, 
namely $(-1,0,\dots,0)$, which makes an angle $3\pi/4$ with the first 
basis vector of $A_{n-1}$, is orthogonal to the rest, and is shorter than the 
rest of the basis vectors by a factor of $1/\sqrt{2}$, {\it i.e.}, it has 
exactly the properties of the ``double bond'' of the Dynkin diagram of $B_n$.  
In other words, $Z_n$ is the root lattice of the Lie group $B_n$.  

The last exercise illustrates the fact that a given root lattice can be 
spanned by different sets of basis vectors (and hence different Lie groups), 
and the choice most convenient for the study of Lie groups is not 
necessarily the most convenient for our purposes.  
As we have seen, the set of nearest neighbors ${\cal S}_\rho$ plays a 
special role in our discussion.  
As a result, we have chosen to define root lattices in a way which depends  
only on the relative directions of the nearest neighbors.  
The unique Lie group defined by the nearest neighbors on a given root 
lattice is always simply laced (no multiple bonds, all basis vectors have 
the same length).  
For $A_n$ it is SU$(n+1)$, for $D_n$ it is SO$(2n)$ and for $Z_n$ it is 
SU$(2)^n$ (not SO$(2n-1)$).\footnote
{This is why we avoid calling the hypercubical lattice $B_n$ --- to emphasize 
that we are studying the root lattice, not the Lie group.}  

The remaining Dynkin diagrams ($C_n$, $F_4$ and $G_2$) are all multiply 
laced (with multiple bonds, and basis vectors of different lengths).  
One can easily check that the $C_n$ and $D_n$ groups have the same root 
lattices, but the former is doubly laced while the latter is simply laced.  
Similarly, the simply laced counterparts of $G_2$ and $F_4$ are $A_2$ and 
$D_4$ respectively\footnote{The authors of Ref.~\cite{Neu1,Neu2,Neu3} 
actually referred to the Celmaster lattice $D_4$ as the $F_4$ lattice.}.  

So we have clarified the relationship between the Lie groups (which have 
little to do with this paper) and the root lattices (our real concern).  
The reader should be warned that, for the purposes of this paper, it is 
advisable to temporarily forget about the Lie group structures of the 
lattices and regard the root lattices as geometrical constructions in their 
own right.  

\section{Discretized Laplacian Operators on Root Lattices}

In this section, we will give explicit formulae for $\Sigma$ and $D^2$ on 
root lattices $Z_n$, $A_n$ and $D_n$.  
Recall that, since all lattices are symmetrical under reflection with 
respect to the origin ($X_m \to -X_m$ for all $m$, $m$ being the 
dimensionality of the coordinate space, {\it i.e.}, $n$ for $Z_n$ and $D_N$ 
and $n+1$ for $A_n$), in the Taylor expansion of $\Sigma$, only terms with 
even powers of $\rho$ survive.  
Hence 
\begin{equation}
\Sigma = {\textstyle{1\over\tau}} \sum_{X_k\in{\cal S}_\rho} \; 
\exp(X_k \cdot \partial) \equiv {\cal O}_0 + a^2 {\cal O}_2 + a^4 {\cal O}_4 + 
a^6 {\cal O}_6 + \dots . 
\end{equation}
The operators ${\cal O}_l$ are $l$-th order derivatives, their forms 
constrained by the symmetries of the lattices.  
As mentioned before, all three root lattices defined in Eq.~(\ref{ZAD}) 
are invariant under permutations of the coordinates.  
Moreover, $Z_n$ and $D_n$ are also symmetric under reflection of individual 
coordinates ($X_m \to -X_m$ for a given $m$).  
These symmetry constraints limit the number of possible operators appearing 
at each order.  

To enumerate the number of possible operators appearing at each order, the 
following definitions are useful.  
\begin{eqnarray}
{\cal B}_{j_1j_2\dots j_p}^{(m)} &=& \sum_{1\le i_1 \neq 
i_2\neq\dots\neq i_p\le m} \partial_{i_1}^{j_1} \partial_{i_2}^{j_2} \dots 
\partial_{i_p}^{j_p}, \quad \hbox{all $j$'s even,}\nonumber\\
{\cal A}_{j_1j_2\dots j_p}^{(m)} &=& \sum_{1\le i_1 \neq 
i_2\neq\dots\neq i_p\le m} \partial_{i_1}^{j_1} \partial_{i_2}^{j_2} \dots 
\partial_{i_p}^{j_p}, \quad \hbox{some $j$'s odd.}
\end{eqnarray}
The definitions of the even operators $\cal B$ and the odd operators 
$\cal A$ are actually identical except that $\cal A$ may contain odd powers 
of $\partial_i$, which is not invariant under reflection of coordinate $x_i$.  
As a result, ${\cal O}_l$ for $Z_n$ and $D_n$ contains only terms like 
${\cal B}_{j_1j_2\dots j_p}^{(n)}$ with $\sum_p j_p = l$ but no $\cal A$ 
terms.  
On the other hand, for $A_n$, both ${\cal B}_{j_1j_2\dots j_p}^{(n+1)}$ and 
${\cal A}_{j_1j_2\dots j_p}^{(n+1)}$ contribute.  

It is easy to see that ${\cal O}_0=1$ for any lattice.  
There are two operators which appear at order $a^2$, namely 
\begin{eqnarray}
{\cal B}_2^{(m)} =& \sum_{1\le i \le m} &\partial_i^2, \nonumber\\
{\cal A}_{11}^{(m)} =& \sum_{1\le i\neq j \le m} &\partial_i \partial_j . 
\end{eqnarray}
As mentioned before, for $Z_n$ and $D_n$, $\cal A$ does not contribute and 
${\cal O}_2$ must be proportional to the even operator ${\cal B}_2^{(m)}$, 
which is exactly the Laplacian operator in $m$ dimensions, and $m=n$ for 
$Z_n$ and $D_n$.  
On the other hand, for $A_n$, ${\cal O}_2$ cannot be simply proportional to 
${\cal B}_2^{(m)}$, as we want something proportional to the Laplacian 
operator in $n$, not $m=n+1$, dimensions.  
However, for $A_n$ the odd operator ${\cal A}_{11}^{(m)}$ does contribute, and 
as we will see below, exactly in the right amount to give the Laplacian 
operator in an $n$ dimensional subspace.  

When one goes to order $a^4$, more operators appear.  
\begin{eqnarray}
{\cal B}_4^{(m)} =& \sum_{1\le i \le m} &\partial_i^4, \nonumber\\
{\cal B}_{22}^{(m)} =& \sum_{1\le i\neq j \le m} &\partial_i^2 \partial_j^2, 
\nonumber\\
{\cal A}_{31}^{(m)} =& \sum_{1\le i\neq j \le m} &\partial_i^3\partial_j, 
\nonumber\\
{\cal A}_{211}^{(m)} =& \sum_{1\le i\neq j\neq k \le m} &\partial_i^2 
\partial_j \partial_k, \nonumber\\
{\cal A}_{1111}^{(m)} =& \sum_{1\le i\neq j\neq k\neq l \le m} &\partial_i 
\partial_j \partial_k \partial_l.
\end{eqnarray}
As before, the odd operators only contribute in the $A_n$ case while the 
even ones in general may contribute for all three classes of root lattices.  

After discussing the generalities, we will move on to specific cases and 
calculate $\Sigma$, and hence $D^2 = 2n(\Sigma-1)/ \rho^2$, for all three 
classes of root lattices.  
We will also list the kissing numbers $\tau$, packing diameters $\rho$ and 
densities of lattice sites $\delta$ of the lattices.  
Many of the properties listed below are extracted from the excellent book 
by Conway and Sloane \cite{Con}, although the definitions of $\rho$ and 
$\delta$ here are different from those in the book.  

\subsection{The Hypercubical Lattice $Z_n$}

The hypercubical lattice $Z_n$, defined as 
\begin{equation}
Z_n \equiv \{X=(x_1,\dots,x_n) \in a{\bf Z}^n \}, 
\end{equation}
has kissing number $\tau=2n$, packing diameter $\rho=a$ and density 
$\delta = 1/a^n$.  
One can easily evaluate $\Sigma$, which takes the form 
\begin{equation}
\Sigma = 1 + \textstyle{1\over2!n} a^2 {\cal B}_2^{(n)} + 
\textstyle{1\over4!n} a^4 {\cal B}_4^{(n)} + {\cal O}(a^6) .
\end{equation}
One can easy show that, due to the orthogonal nature of $Z_n$, 
\begin{equation}
{\cal O}_l = {\textstyle{1\over l!n}}\sum_{1\le i \le n}\partial_i^l . 
\end{equation}
Lastly, the one shell discretization $D^2$ is 
\begin{equation}
D^2 = {\cal B}_2^{(n)} + {\textstyle{2\over 4!}} a^2 {\cal B}_4^{(n)} + \dots 
= \partial^2 + {\textstyle{1\over 12}} a^2 \sum_{1\leq i\leq n} \partial_i^4
+ {\cal O}(a^4), 
\end{equation}
and as predicted, $D^2 = \partial^2$ in the leading order.  

\subsection{The Checkerboard Lattice $D_n$}

The checkerboard lattice $D_n$, defined as 
\begin{equation}
D_n \equiv \{X=(x_1,\dots,x_n) \in a{\bf Z}^n; \sum_k x_k \in 2a{\bf Z}\},
\end{equation}
has kissing number $\tau=2n(n-1)$ and packing diameter $\rho = \sqrt{2}a$.  
Only half of the lattice sites of $Z_n$ is occupied in $D_n$, so its 
density $\delta = 1/2a^n$ is half of that of $Z_n$.  
It is straightforward to calculate $\Sigma$,  
\begin{equation}
\Sigma = 1 + \textstyle{2\over 2!n} a^2 {\cal B}_2^{(n)} + 
\textstyle{2\over 4!n} a^4 {\cal B}_4^{(n)} + 
\textstyle{6\over 4!n(n-1)} a^4 {\cal B}_{22}^{(n)} + {\cal O}(a^6),
\end{equation}
and bearing in mind that now $\rho^2 = 2 a^2$, 
\begin{eqnarray}
D^2 &=& {\cal B}_2^{(n)} + {\textstyle{2\over 4!}} a^2 {\cal B}_4^{(n)} + 
{\textstyle{6\over 4!(n-1)}} a^2 {\cal B}_{22}^{(n)} + {\cal O}(a^4)\nonumber\\
&=& \partial^2 + {\textstyle{1\over 12}} a^2 \sum_{1\leq i\leq n} \partial_i^4
+ {\textstyle{1\over 4(n-1)}} a^2 \sum_{1\leq i\neq j\leq n} \partial_i^2 
\partial_j^2 + {\cal O}(a^4).  
\end{eqnarray}
Again, we verify that $D^2 = \partial^2$ in the leading order\footnote{
This discretization is equivalent to that used in Ref.~\cite{Neu1,Neu2,Neu3}}.

The four dimensional checkerboard lattice $D_4$ is of special importance 
as it has been extensively studied by Celmaster \cite{Cel1,Cel2,Cel3} and 
is usually known as the Celmaster, or body-centered hypercubical lattice.  
\begin{equation}
D^*_4 \equiv \{X=(x_1,x_2,x_3,x_4) \in a{\bf Z}^4 \; \cup \; 
a({\bf Z}^4 + (\textstyle{1\over2}, {1\over2}, {1\over2}, {1\over2})) \},
\end{equation}
{\it i.e.}, the coordinates are either all integers or all half integers.  
The Celmaster lattice $D^*_4$ is actually the dual, or reciprocal lattice, 
of $D_4$.  
Reciprocal lattices are in general quite different: in three dimensions, 
the face-centered cubic and body-centered cubic lattices are reciprocal, 
and are not similar at all.  
In this special case, however, $D_4$ and $D^*_4$ are actually similar to 
each other.  
First note that both have 24 sites in the first shell: 
\begin{eqnarray}
D_4&:& \quad (\pm1,\pm1,0,0) + \hbox{perm.}; \nonumber\\
D^*_4&:& \quad (\pm1,0,0,0) + \hbox{perm.}, \qquad (\textstyle
{\pm{1\over2},\pm{1\over2},\pm{1\over2},\pm{1\over2}}) + \hbox{perm.};
\label{D1}
\end{eqnarray}
and 24 sites on the second shell 
\begin{eqnarray}
D_4&:& \quad (\pm2,0,0,0) + \hbox{perm.}, \qquad (\pm1,\pm1,\pm1,\pm1) + 
\hbox{perm.}; \label{D2}\\ D^*_4&:& \quad (\pm1,\pm1,0,0) + \hbox{perm.}\quad. 
\nonumber
\end{eqnarray}
It is clear that the first shell of $D_4$ is exactly the second shell 
of $D^*_4$ and the second shell of $D_4$ is the first shell of $D^*_4$ 
scaled up by a factor of 2.  
One can actually show that one can start out with $D_4$, perform a global 
rotation and overall rescaling by a factor of $1/\sqrt{2}$ and end up with 
$D^*_4$.  
So $D_4$ and $D^*_4$ are geometrically similar in the sense that similar 
triangles are similar --- they can be mapped onto each other through 
global rotation and overall scaling.  
As we will see below, the first and second shell of $D_4$ being similar 
will be important when we try to cancel ${\cal O}(a^2)$ errors by going 
beyond one shell discretization.  

\subsection{The Simplicial Lattice $A_n$}

Lastly, we will discuss the simplicial lattice $A_n$, which is the hardest 
to study as it is embedded in an $n+1$-dimensional coordinate space.  
\begin{equation}
A_n \equiv \{X=(x_1,\dots,x_n,x_{n+1})\;\in a{\bf Z}^{n+1}; \sum_k x_k = 0\}.
\end{equation}
it has kissing number $\tau=n(n+1)$, packing diameter $\rho = \sqrt{2}a$, 
and density\footnote{
The easiest way to show this is by observing that $Z_{n+1}$ has density 
$1/a^{n+1}$, and the $A_n$-like sublattices $\{X \in a{\bf Z}^{n+1}; 
\sum_k x_k = C\}$ where $C$ is a fixed integer, have density $\sqrt{n+1}/a$ 
in $Z_{n+1}$.  
The quotient of the two gives the density of $A_n$.} 
$\delta = 1/\sqrt{n+1}a^n$, and 
\begin{eqnarray}
\Sigma &=& 1 + \textstyle{2\over 2!(n+1)} a^2 {\cal B}_2^{(n+1)} + 
\textstyle{-2\over 2!n(n+1)} a^2 {\cal A}_{11}^{(n+1)} \nonumber\\ & &\quad +
\textstyle{2\over 4!(n+1)} a^4 {\cal B}_4^{(n+1)} + 
\textstyle{6\over 4!n(n+1)} a^4 {\cal B}_{22}^{(n+1)} + 
\textstyle{-6\over 4!n(n+1)} a^4 {\cal A}_{31}^{(n+1)} + {\cal O}(a^6),
\end{eqnarray}
without any terms proportional to ${\cal A}_{211}$ or ${\cal A}_{1111}$.  
One can then read off $D^2$, which takes a rather complicated form.  
\begin{eqnarray}
D^2 &=& \textstyle{n\over (n+1)} {\cal B}_2^{(n+1)} - 
\textstyle{1\over (n+1)} {\cal A}_{11}^{(n+1)} \nonumber \\& &\quad + 
\textstyle{2n\over 4!(n+1)} a^2 {\cal B}_4^{(n+1)} + 
\textstyle{6n\over 4!n(n+1)} a^2 {\cal B}_{22}^{(n+1)} - 
\textstyle{6n\over 4!n(n+1)} a^2 {\cal A}_{31}^{(n+1)} + {\cal O}(a^4)\\
&=& {\textstyle{n\over (n+1)}} \sum_{1\le i \le n+1} \partial_i^2 - 
{\textstyle{1\over (n+1)}} \sum_{1\le i \neq j\le n+1} \partial_i\partial_j 
\nonumber\\ & &\quad + {\textstyle{n\over 12(n+1)}} a^2 \sum_{1\le i \le n+1} 
\partial_i^4 + {\textstyle{1\over 4(n+1)}} a^2 \sum_{1\le i \neq j\le n+1} 
\partial_i^2\partial_j^2 - \textstyle{{1\over 4(n+1)}} a^2 
\sum_{1\le i \neq j\le n+1} \partial_i^3\partial_j^1 + {\cal O}(a^4). 
\nonumber
\end{eqnarray}
Does the sum of the first two terms give the Laplacian operator in the 
$n$-dimensional subspace described by $\sum x_i = 0$?  
Note the following relation between the Laplacian operators in $n$ and $n+1$ 
dimensions.  
\begin{equation}
\partial^2\Big|_{{\bf R}^{n+1}} = \partial^2\Big|_{{\bf R}^n} + 
\partial_{\bar x}^2, 
\end{equation}
where $\bar x=(1,1,\dots,1)/\sqrt{n+1}$ is the unit normal vector of the 
subspace $\sum x_i = 0$.  
One can easily check that 
\begin{equation}
\partial_{\bar x}^2 = \bigg( \sum_{1\le i \le n+1} \partial_i^2 
+ \sum_{1\le i \neq j\le n+1} \partial_i\partial_j \bigg) \bigg/(n+1), 
\end{equation}
and hence 
\begin{equation}
\partial^2\Big|_{{\bf R}^n} = \partial^2\Big|_{{\bf R}^{n+1}} - 
\partial_{\bar x}^2 = {\textstyle{n\over (n+1)}} \sum_{1\le i \le n+1} 
\partial_i^2 - {\textstyle{1\over (n+1)}} \sum_{1\le i \neq j\le n+1} 
\partial_i\partial_j ,
\end{equation}
exactly as advertised.  

\section{Discretization Errors} 

We have seen that $D^2$ does reproduce $\partial^2$ exactly in leading order 
for all three classes of root lattices we have studied.  
In this section, we will analyze the order $a^2$ errors of $D^2$, which 
we will denote by $E$ and which is related to ${\cal O}_4$ by 
\begin{equation}
E = 2n {\cal O}_4 a^4/ \rho^2.  
\end{equation}
Its forms for all three classes of root lattices are summarized below.  
\begin{eqnarray}
Z_n &: E= {\textstyle{1\over 4!}} a^2 &2 {\cal B}_4^{(n)} , \nonumber\\
D_n &: E= {\textstyle{1\over 4!}} a^2 &\big(2 {\cal B}_4^{(n)} 
+ \textstyle{6\over n-1}  {\cal B}_{22}^{(n)} \big) , \\
A_n &: E= {\textstyle{1\over 4!}} a^2 &\big(\textstyle{2n\over n+1} 
{\cal B}_4^{(n+1)} + \textstyle{6n\over n(n+1)} {\cal B}_{22}^{(n+1)} - 
\textstyle{6n\over n(n+1)} {\cal A}_{31}^{(n+1)} \big) . \nonumber
\end{eqnarray}
At first sight the coefficients may look random and without any discernible 
patterns, but as we will see below they are actually not as random as they 
look.  

All the expressions for $E$ above are quartic polynomials of $\partial_i$.  
Naively, one may estimate their size by $p^4$, where $p$ is the typical 
momentum scale in question.  
However, such an estimate is sloppy, since as an operator $p^4 = \partial^4 
= (\partial^2)^2$, {\it i.e.}, the square of the Laplacian operator, which 
is a scalar operator in $n$-dimensional Euclidean space, while ${\cal B}_4$, 
${\cal B}_{22}$ and ${\cal A}_{31}$ are all non-scalar operators as defined.  
Thus motivated, we will define $E_0$, the scalar or isotropic part of $E$, as 
\begin{equation}
E_0 = \int d \mu_{n-1}\;E =\;\hbox{(dimensionless numerical factor)}\;a^2 
\partial^4, 
\end{equation}
where $d\mu_{n-1}$ is again the unit measure on $S^{n-1}$.  
In other words, $E_0$ is $E$ integrated over all directions in $n$-dimensional 
momentum space.  
When one evaluates $E_0$, it turns out that 
\begin{equation}
E_0 = {\textstyle{1\over 4!}} \partial^4 a^2\times\cases{6/(n+2), &for $Z_n$;
\cr 12/(n+2), &for $D_n$ or $A_n$, \cr}
\end{equation}
which is surprisingly simple (when contrasted with the expressions before 
the integration).  
Moreover, for all cases one has  
\begin{equation}
E_0 = \rho^2 \partial^4/4(n+2).
\label{E0}
\end{equation}

The above relation is not obvious (at least to the author) from the raw 
expressions for $E$, but once motivated it can actually be proved very 
easily for all lattices (not necessarily root lattices).  
Recall that $\Sigma$ is defined as 
\begin{equation}
\Sigma = \bigg\{ {\textstyle{1\over\tau}} \sum_{X_k\in{\cal S}_\rho} 
\; \exp(X_k \cdot \partial) \bigg\}.  
\end{equation}
For each site $X_k\in{\cal S}_\rho$, one can expand the exponential as 
\begin{equation}
\exp(X_k \cdot \partial) = 1 + \textstyle{1\over2!} (X_k \cdot \partial)^2 
+ \textstyle{1\over4!} (X_k \cdot \partial)^4 + \;\hbox{odd terms}\;+ 
{\cal O}(\rho^6), 
\end{equation}
where the odd terms will eventually be cancelled and do not concern us here.  
The quadratic and quartic terms can be rearranged in the following way. 
\begin{eqnarray}
(X_k \cdot \partial)^2 &=& \rho^2 \partial^2 /n + 
\Big[(X_k \cdot \partial)^2 -\rho^2 \partial^2 /n\Big], \nonumber\\
(X_k \cdot \partial)^4 &=& 3 \rho^4 \partial^4 /n(n+2) + 
\Big[(X_k \cdot \partial)^4 - 3 \rho^4 \partial^4 /n(n+2)\Big].  
\label{Sym}
\end{eqnarray}
The integrals 
\begin{equation}
\int d\mu_{n-1} \; (X_k \cdot \partial)^2 = \rho^2 \partial^2 /n, \qquad 
\int d\mu_{n-1} \; (X_k \cdot \partial)^4 = 3 \rho^2 \partial^4 /n(n+2), 
\end{equation}
decree that the terms in square brackets in Eqs.~(\ref{Sym}) vanish under 
integration over all directions.  
Hence 
\begin{equation}
\int d\mu_{n-1} \; \Sigma = {\textstyle{1\over\tau}} 
\sum_{X_k\in{\cal S}_\rho} \big(1 + \textstyle{1\over2!n} \rho^2 \partial^2 + 
\textstyle{3\over4!n(n+2)} \rho^4 \partial^4 \big) + {\cal O}(\rho^6) , 
\end{equation}
which is a sum over $\tau$ identical expressions and then divided by $\tau$, 
and is equal to the expression itself, 
\begin{equation}
\int d\mu_{n-1} \; \Sigma = 1 + \textstyle{1\over2!n} \rho^2 \partial^2 + 
\textstyle{3\over4!n(n+2)} \rho^4 \partial^4 + {\cal O}(\rho^6).  
\end{equation}
Lastly, we have the scalar part of $D^2$.  
\begin{equation}
\int d\mu_{n-1} \; D^2 = \int d\mu_{n-1} \; 2n (\Sigma - 1) /\rho^2 = 
\partial^2 + \textstyle{6\over4!(n+2)} \rho^2 \partial^4 + {\cal O}(\rho^4), 
\end{equation}
and $E_0$ can be read off simply as the $\rho^2$ term.  

The simple result Eq.~(\ref{E0}) has important implications.  
Firstly, it gives us the following {\it impossibility theorem}: 

\bigskip

{\sl One shell discretization of the Laplacian operator always suffers 
from order $a^2$ errors, regardless of the choice of the lattice.}  

\bigskip

\noindent One might have hope that, by using a lattice with a high degree of 
symmetry, all ${\cal O}(a^2)$ terms may vanish identically.  
We now see that is an impossible dream.  
There is always a scalar part which is not prohibited by any symmetry.  
The only way to eliminate order $a^2$ errors is by involving lattice sites 
which are not nearest neighbors, as we will discuss later in this paper.  

Another corollary of Eq.~(\ref{E0}) is that the scalar discretization error
$E_0$ depends only on a single geometrical parameter --- the packing diameter 
$\rho$.  
(It also depends on the spacetime dimensionality $n$, but that is usually 
determined by the physics problem in question.) 
One can then ask, in a given spacetime dimension, which lattice will 
minimize $E_0$?  
It will be the subject of the next section.  

\section{Error Densities of One Shell Discretizations}

As defined, the scalar discretization error $E_0$ is independent of how 
we describe the lattice on the coordinate space.  
This is certainly a pleasing feature to physicists, who know that all 
physical quantities should not depend on the choice of coordinates.  
On the other hand, $E_0$ is proportional to $\rho^2$, which is not surprising; 
the smaller $\rho^2$, the more lattice sites we are using to describe a 
given physical volume, and hence the smaller is the error.  
In other words, the error grows with $1/\delta$, the volume of the unitcell 
of the lattice.  
To determine which class ($Z_n$, $D_n$ or $A_n$) is optimal for our purposes, 
one defines the {\it error density\/} $\epsilon$ as the ratio of the $E_0$ 
to (a certain power of) the unitcell volume.  
\begin{equation}
E_0 = {1\over 4(n+2)} \, \epsilon \, \delta^{-2/n} \, \partial^4.  
\end{equation}
The $-2/n$ power has been chosen such that $\epsilon$ is $\rho$ independent; 
recall that $E_0$ and $\delta$ scale like $\rho^2$ and $\rho^{-n}$ 
respectively.  
Hence $\epsilon$ thus defined is a sheer parameter (not an operator) which 
just depends on which lattice which are using.  
For root lattices, it carries the value 
\begin{equation}
\epsilon = \cases{1,& for $Z_n$; \cr 2^{1-2/n},& for 
$D_n$; \cr 2(n+1)^{-1/n},& for $A_n$. \cr}
\end{equation}
Obviously one can check that the error density of $Z_n$ is always smaller than 
that of $D_n$ and $A_n$.\footnote{
Moreover, one also has $\epsilon_{Z_2} = \epsilon_{D_2}$ and $\epsilon_{A_3} 
= \epsilon_{D_3}$, {\it i.e.}, the error density of a lattice is independent 
of how it is named.  }
This is certainly not a surprising result, as we have shown in the previous 
section that $E_0$ is proportional just to $\rho^2$, and for our root 
lattices with the same $\rho$, $Z_n$ is the one with the smallest density.  

We have seen that, {\it among root lattices}, $Z_n$ has the smallest error 
density.  
We can actually prove the much stronger result that $Z_n$ has the 
smallest error density among all lattices whose ${\cal S}_\rho$ have 
vanishing quadrapole moment, {\it i.e.}, all the lattices with 
$D^2 = \partial^2$ in leading order.  
The proof runs in two steps.  
First, observe that among all lattices whose basis vectors can be chosen 
to be nearest neighbors, $Z_n$ has the lowest density.  
(Among all parallelohedra with all sides of unit length, the one with the 
largest volume is a cube.)  
So $Z_n$ has the lowest error density among them.  
What about lattices whose basis vectors {\it cannot\/} all be chosen as 
nearest neighbors?  
Note that, for such a lattice $\Lambda$, there exists a proper sublattice 
$\bar \Lambda \subset \Lambda$ which is generated by the nearest neighbors 
of $\Lambda$.  
The lattices $\Lambda$ and $\bar \Lambda$ have the same $\rho$ and hence 
the same $E_0$, yet being a subset, the density of $\bar\Lambda$ is smaller 
than that of $\Lambda$.
As a result, the error density of $\Lambda$ is larger than that of $\bar 
\Lambda$, which in turn by the previous argument is larger than that of 
$Z_n$.  
So we have the important result: 

\bigskip

{\sl Under one shell discretization, the hypercubical lattices have the 
smallest error densities among all possible choices of lattices in the 
same number of spacetime dimensions.}

\bigskip

Most lattice simulations are performed on hypercubical lattices, but usually 
the reasons are historical (following Wilson's example in Ref.~\cite{Wil}) or 
algorithmic ($D^2$ can be easily written down as in Eq.~(\ref{Z}) by using 
the trapezoidal rule).  
Here we provide another rationale: the hypercubical lattices have the lowest 
error density.  

\section{Anisotropy}

While $E_0$ measures the isotropic part of the order $a^2$ errors, $E-E_0$ 
measures the anisotropy of $D^2$. 
The operator $E-E_0$ is a quartic polynomial in $\partial_i$, which in 
general can have all $\ell$-th multipole moment up to $\ell=4$.  
However, the first and third (dipole and octupole) multipole moments are 
identically zero  as demanded by reflection symmetry, the second (quadrapole) 
moment is explicitly required to vanish, and the zeroth (monopole) moment 
$E_0$ has been subtracted out.  
So $E-E_0$ is purely of the fourth (hecadexapole) moment.  

It is straightforward to verify that $E-E_0$ is non-vanishing for all root 
lattices with one exception: $D_4$, the checkerboard lattice in four 
dimensions. 
\begin{equation}
E= {\textstyle{1\over 4!}} 2 a^2 \big({\cal B}_4 + {\cal B}_{22} \big) 
= {\textstyle{2\over 4!}} a^2 \partial^4 = E_0 \qquad \hbox{for $D_4$.}
\end{equation}
That $D_4$ is more isotropic than $Z_4$ has been noted by Celmaster in 
Ref.~\cite{Cel1}, but here the result is stronger.  

\bigskip

{\sl The checkerboard lattice in four dimensions $D_4$ is exactly isotropic 
at order $a^2$.  
It is the only unexceptional root lattice with this property.}

\bigskip

\noindent One can understand the isotropy of the $D_4$ lattice by studying 
its geometry.  
The $D_4$ lattice has an accidental threefold discrete symmetry (which is 
also a symmetry of its Dynkin diagram) which mixes ${\cal B}_4$ and 
${\cal B}_{22}$.  
The only combination which is invariant under this threefold symmetry is 
$\partial^4$.  
In other words, rotational symmetry is protected by this accidental 
threefold discrete symmetry at order $a^2$.  

Since we are mainly interested in lattices in low dimensions ($n\leq 4$), 
we have not studied the anisotropy of $E_{6,7,8}$, although the 
calculation should be straightforward (though tedious --- the packing 
number $\tau$ is 240 for $E_8$).  
We also have not studied the issue of anisotropy for non-root lattices.  
However, since we have seen that the isotropy of $D_4$ is related to an 
accidental discrete symmetry, a feature not apparent in these other cases, 
the author would be very surprised if other lattices share this property 
of isotropy.  

\section{Beyond One Shell Discretization}

We have seen that order $a^2$ errors are unavoidable in one shell 
discretization.  
The order $a^2$ errors can be substantial on coarse lattices; for example, 
as a rule of thumb the order $a^2$ error on a hypercubical lattice is around 
10\% if $a$ is chosen to be around a third of the size of the physical 
system in question \cite{GPL}.  
In order to reduce the error, one may try to use finer lattices.  
It is, however, a very costly procedure as the cost of lattice simulations 
grows at least as fast as the number of lattice sites, which is proportional 
to $a^{-n}$.  
(And that is before taking critical slowing down into account.)  
For $n=4$, to reduce the error to 1\% would increase the cost a hundredfold.  
Another way of seeing the same result is by observing that the error density 
$\epsilon$, which is $a$ independent, is the proportionality constant 
between $E_0$ and $\delta^{-2/n}$.  
To reduce $E_0$ by a factor of 10, $\delta$ must be decreased by a factor 
of $10^{n/2}$, {\it i.e.}, a hundredfold in four dimensions.  

Clearly it is not practical to control the $a^2$ errors just by using finer 
and finer lattices. 
Instead, one can completely eliminate the $a^2$ error by involving sites 
not in ${\cal S}_\rho$, {\it i.e.}, by going beyond one shell discretization.  
As an example, we will review how this is done on hypercubical lattices.  

Recall that, for hypercubical lattices, one shell discretization on 
${\cal S}_a$ gives 
\begin{equation}
D^2[{\cal S}_a] = {\cal B}_2^{(n)} + {\textstyle{2\over 4!}} a^2 
{\cal B}_4^{(n)} + {\cal O}(a^4)  = \partial^2 + {\textstyle{1\over 12}} a^2 
\sum_{1\leq i\leq n} \partial_i^4 + {\cal O}(a^4), 
\end{equation}
The argument in the square brackets is to remind us of which shell we are 
using.  
One can also discretize on the following set of $2n$ sites:
\begin{equation}
2{\cal S}_a = \{ (\pm 2a,0,\dots,0) + \; \hbox{perm.} \} .  
\end{equation}
Notice that $2{\cal S}_a$ is just ${\cal S}_a$ scaled up by a factor of 2. 
As a result, 
\begin{equation}
D^2[2{\cal S}_a] = \partial^2 + {\textstyle{1\over 12}} (2a)^2 
\sum_{1\leq i\leq n} \partial_i^4 + {\cal O}(a^4).   
\end{equation}
Consequently, the combination ${4\over 3} D^2[{\cal S}_a] - {1\over 3} 
D^2[2{\cal S}_a]$ is free of order $a^2$ errors.   

Notice that $2{\cal S}_a$ is {\it not\/} the second shell of $Z_n$.  
(In fact it is just part of the fourth shell.)  
Since the errors are in general proportional to some high powers of the 
radii of the shells involved, can we do better by using just the first and 
second shell, ${\cal S}_a$ and ${\cal S}_{\sqrt{2}a}$?  
Recall that ${\cal S}_{\sqrt{2}a}$ is the first shell of $D_n$, and $D^2$ 
has been calculated to be 
\begin{equation}
D^2[{\cal S}_{\sqrt{2}a}] = {\cal B}_2^{(n)} + {\textstyle{2\over 4!}} a^2 
{\cal B}_4^{(n)} + {\textstyle{6\over 4!(n-1)}} a^2 {\cal B}_{22}^{(n)} 
+ {\cal O}(a^4) , 
\end{equation}
which contains both ${\cal B}_4^{(n)}$ and ${\cal B}_{22}^{(n)}$.  
In contrast, $D^2[{\cal S}_a]$ involves just the former but not the latter.  
Thus one cannot obtain an expression free of both ${\cal B}_4^{(n)}$ and 
${\cal B}_{22}^{(n)}$ errors by taking a linear combination of 
$D^2[{\cal S}_a]$ and $D^2[{\cal S}_{\sqrt{2}a}]$.  
In other words, one cannot eliminate the order $a^2$ order by taking a linear 
combination of $D^2$ of any two shells.  

Bearing the above examples on the hypercubical lattices in mind, we will 
now move on to discuss the framework of elimination of the order $a^2$ errors 
on lattices in general.  
Consider the one shell discretization on two shells of radii $r_{1,2}$ 
with $r_1<r_2$.  
Then we want to see if the expression 
\begin{equation}
D^2[{\cal S}_{r_1},{\cal S}_{r_2}] = c_1 D^2[{\cal S}_{r_1}] - c_2 
D^2[{\cal S}_{r_2}], 
\end{equation}
is equal to the Laplacian operator without any $a^2$ errors with some choices 
of coefficients $c_{1,2}$.  
Obviously in order to get the coefficient of the Laplacian operator to be 
unity we need $c_1 - c_2 = 1$.  
We also want the scalar errors $E_0$ of the two shells, which are 
proportional to $r^2$, to cancel, by demanding $c_1 r_1^2 - c_2 r_2^2 =0$, 
or $c_1/c_2 = (r_1/r_2)^{-2}$. 
Putting these together we have 
\begin{equation}
c_1 = r_2^2/(r_2^2 - r_1^2), \qquad c_2 = r_1^2/(r_2^2 - r_1^2). 
\end{equation}

So now we can define the {\it two shell discretization\/} of the 
Laplacian operator over shells ${\cal S}_{r_1}$ and ${\cal S}_{r_2}$ as 
\begin{eqnarray}
D^2[{\cal S}_{r_1},{\cal S}_{r_2}] &=& \bigg(r_2^2 D^2[{\cal S}_{r_1}] 
- r_1^2 D^2[{\cal S}_{r_2}] \bigg)\bigg/(r_2^2 - r_1^2) \nonumber\\ 
&\equiv& \partial^2 + a^2 {\cal H}_2 + a^4 {\cal H}_4 + {\cal O}(a^6).  
\end{eqnarray}
Thus defined, the scalar discretization error of order $a^2$ is guaranteed 
to vanish, {\it i.e.}, $\int d\mu_{n-1} \; {\cal H}_2 = 0$.  
However, ${\cal H}_2$ itself may be non-zero, as shown by the example with 
the first and second shell in $Z_n$.  

When we defined one shell discretization early in this paper, we had to 
explicitly demand that the shell in question have zero quadrapole moment.  
Here the situation is similar.  
For two shell discretization we have to demand that the two shell 
configuration in question, after weighting by $c_{1,2}$, has vanishing 
hexadecapole moment.  
In other words, the hexadecapole moments of the two shells must cancel 
each other.  

In principle, one can translate the condition of the cancellation of 
hexadecapole moments into mathematical constraints and look for their most 
general solutions --- but in practice it is almost impossible.  
(We have not even done that for the quadrapole case!)  
For example, for $Z_n$ or $D_n$, the shells described by $(y_1, \dots y_n)$ 
and $(z_1, \dots z_n)$ (plus permutation) will have their fourth multipole 
moment cancelled if and only if 
\begin{equation}
\sum_i y_i^4 = K \sum_i z_i^4, \quad \hbox{and} \quad 
\sum_{i,j} y_i^2 y_j^2 = K \sum_{i,j} z_i^2 z_j^2 \qquad \hbox{for some $K$.} 
\end{equation}
These are coupled diophantine-like constraints, which in general cannot be 
solved.  
The corresponding constraints for $A_n$ are even more hideous.  

However, there are special cases which we know the hexadecapole moment 
are guaranteed to cancel, namely when the two shells are identical up to 
scaling (but no rotation).  
To be exact, let's define, for any natural number $p>1$, 
\begin{equation}
p{\cal S}_\rho = \{pX: X\in {\cal S}_\rho\}.  
\end{equation}
Then the two shell discretization $D^2[{\cal S}_\rho,p{\cal S}_\rho]$ has 
vanishing hexadecapole moment and hence is free of order $a^2$ errors.  
The proof is immediate. 
For $D^2[{\cal S}_\rho] = \partial^2 + \rho^2 {\cal H}_2 + {\cal O}(a^4)$, 
one has 
\begin{eqnarray}
D^2[{\cal S}_\rho,p{\cal S}_\rho] &=& \bigg((p\rho)^2 D^2[{\cal S}_\rho] 
- \rho^2 D^2[p{\cal S}_\rho] \bigg)\bigg/ ((p\rho)^2 - \rho^2)\nonumber\\
&=& \bigg(p^2\rho^2 (\partial^2 + \rho^2 {\cal H}_2) - \rho^2 (\partial^2 
+ p^2 \rho^2 {\cal H}_2) \bigg)\bigg/ (p^2-1)\rho^2 + {\cal O}(\rho^4) 
\nonumber\\&=& \partial^2 + {\cal O}(\rho^4).  
\end{eqnarray}

In practice, the two shells should be chosen as small as possible to 
minimize the error. 
This suggests setting $p=2$, and 
\begin{equation}
D^2[{\cal S}_\rho,2{\cal S}_\rho] = \textstyle{4\over3} D^2[{\cal S}_\rho] 
- \textstyle{1\over3} D^2[2{\cal S}_\rho].  
\end{equation}
 
On the other hand, there is a unique case where the hexadecapole moments 
do cancel for two shells which are not related simply by scaling.  
As the reader may have guessed, this is the case for the first and second 
shells of the four dimensional checkerboard lattices, where the hexadecapole 
moment of each shell vanishes identically as dictated by the threefold 
discrete symmetry.  
Recall that both the first shell ${\cal S}_{\sqrt{2}a}$ and the second shell 
${\cal S}_{2a}$ contain 24 sites as described in Eqs.~(\ref{D1},\ref{D2}), and 
the two shell discretization over these two shells is
\begin{equation}
D^2[{\cal S}_{\sqrt{2}a},{\cal S}_{2a}] = 2 D^2[{\cal S}_{\sqrt{2}a}] - 
D^2[{\cal S}_{2a}] \qquad \hbox{for $D_4$}, 
\end{equation}
which can be checked to be free of order $a^2$ errors.  

\section{Error Densities of Two Shell Discretizations}

Consider a two shell discretization with $a^2$ errors completely cancelled.  
\begin{equation}
D^2[{\cal S}_{r_1},{\cal S}_{r_2}] = \partial^2 + a^4 {\cal H}_4 + \dots 
\end{equation}
To study the order $a^4$ errors, one needs to know ${\cal H}_4$, which in 
general is a sextic polynomial of $\partial_i$.  
To calculate ${\cal H}_4$ would be a painful exercise.  
However, it is easy to evaluate $E'_0$, the scalar part of ${\cal H}_4$, 
\begin{equation}
E'_0 = \int d\mu_{n-1} \; a^4 {\cal H}_4 .  
\end{equation}
Again we will start with the Taylor expansion of $\exp(X_k \cdot \partial)$.  
\begin{equation}
\exp(X_k \cdot \partial) = 1 + \textstyle{1\over2!} (X_k \cdot \partial)^2 
+ \textstyle{1\over4!} (X_k \cdot \partial)^4 + \textstyle{1\over6!} 
(X_k \cdot \partial)^6 + \hbox{odd terms} + {\cal O}(a^8).  
\end{equation}
With $X_k \in {\cal S}_r$ (not necessarily the first shell), it can be 
rewritten as 
\begin{eqnarray}
\exp(X_k \cdot \partial) &=& 1 + \textstyle{1\over2!} I_2 r^2 \partial^2 
+ \textstyle{1\over4!} I_4 r^4 \partial^4 + \textstyle{1\over6!} I_6 r^6 
\partial^6 \nonumber\\ & &\quad + \; [\dots] + \hbox{odd terms} + {\cal O}(r^8), 
\label{E2}
\end{eqnarray}
where $I_l$ is defined by 
\begin{equation}
\int d\mu_{n-1} (X_k \cdot \partial)^l = I_l r^l \partial^l, 
\end{equation}
with $I_2 = 1/n$ and $I_4 = 3/n(n+2)$.  
We have not calculated $I_6$ which fortunately is not needed for our 
purposes.  
Lastly, the $[\dots]$ in Eq.~(\ref{E2}), like the terms in square brackets in 
Eq.~(\ref{Sym}), vanishes under integration over all directions.  

The scalar part of the one shell discretization on ${\cal S}_r$ is 
\begin{equation}
\int d\mu_{n-1} \; D^2[{\cal S}_r] = \partial^2 + \textstyle{2!\over4!} 
(I_4/I_2) r^2 \partial^4 + \textstyle{2!\over6!} (I_6/I_2) r^4 \partial^6 
+ {\cal O}(r^6).  
\end{equation}
and the scalar part of the two shell discretization on ${\cal S}_{r_1}$ 
and ${\cal S}_{r_2}$ is 
\begin{equation}
D^2[{\cal S}_{r_1},{\cal S}_{r_2}] = \partial^2 - \textstyle{2!\over6!} 
(I_6/I_2) r_1^2 r_2^2 \partial^6 + {\cal O}(r^6), 
\end{equation}
and hence 
\begin{equation}
E'_0 = - \textstyle{2!\over6!} \, (I_6/I_2) \, r_1^2 r_2^2 \, \partial^6 .  
\end{equation}

Now one can define the error density $\epsilon'$ of a two shell 
discretization as 
\begin{equation}
E'_0 = - \textstyle{2!\over6!} \, (I_6/I_2) \, \epsilon' \, \delta^{-4/n} 
\, \partial^6 .  
\end{equation}
As expected, using larger shells will increase the error density.  
For root lattices, and using shells ${\cal S}_\rho$ and $2{\cal S}_\rho$, 
the error densities are 
\begin{equation}
\epsilon' = \cases{4,& for $Z_n$; \cr 2^{4-4/n},& for 
$D_n$; \cr 16(n+1)^{-2/n},& for $A_n$. \cr}
\end{equation}
Again, the error densities are the smallest for hypercubical lattices. 
The reason is the same as the one shell case: for the same $\rho$, $Z_n$ 
has the lowest density.  
And the result can be as easily generalized to non-root lattices as before.  

On the other hand, for $D_4$ we can simply use the first and second shells, 
and the error density is 
\begin{equation}
\epsilon = 4, \quad \hbox{for} \quad D^2[{\cal S}_{\sqrt{2}a},{\cal S}_{2a}], 
\end{equation}
just as low as that of $Z_n$.  
Hence we come to the following {\it conjecture}: 

\bigskip

{\sl Under two shell discretization, the hypercubical lattices have the 
smallest error densities among all possible choices of lattices in the 
same number of spacetime dimensions.
The only other case with an error density as low is two shell discretization 
on the first and second shells of the four dimensional checkerboard lattice.}

\bigskip

\noindent The statement above is just a conjecture, not a theorem, as there 
may be two shells configurations other than ${\cal S}_\rho$ and 
$2{\cal S}_\rho$ which have even lower error densities, just as the $D_4$ 
case we have seen.  
However, exact cancellations of the hexadecapole moments of two shells 
are very unlikely to happen without a symmetry rationale (like the threefold 
discrete symmetry of $D_4$), and since we do not see such extra symmetries 
in other lattices, the author concludes that it is very unlikely that
counterexamples to the above conjecture will be found.  

\section{Conclusion}

Let's recapitulate what we have learned through this study: 

$\bullet$ One shell discretization faithfully reproduces the Laplacian 
operator (with order $a^2$ errors) if and only if the shell has vanishing 
quadrapole moment.  

$\bullet$ The order $a^2$ error for one shell discretization is non-zero 
for any choice of lattice.  
The hypercubical lattices have the smallest error densities in any spacetime 
dimensions.  

$\bullet$ Due to its threefold discrete symmetry, the four dimensional 
checkerboard lattice is isotropic up to order $a^2$.  
It is the only root lattice, and also probably the only lattice (root or 
non-root), with this property.  

$\bullet$ Two shell discretization reproduces the Laplacian operator without 
order $a^2$ errors if and only if the hexadecapole moments of the two shells 
cancel each other. 

$\bullet$ The order $a^2$ error for one shell discretization is non-zero 
for any choice of lattice.  
It is likely (though not yet proved) that the hypercubical lattices have 
the smallest error densities in any spacetime dimensions.  
In four dimensions, the error density of the checkerboard lattice is just 
as small as the hypercubical.  

One can see a clear pattern recurring throughout this discussion.  
The scalar (isotropic) properties of discretizations can be very cleanly 
studied by Taylor expanding $\exp(X_k \cdot \partial)$ and then integrating 
over all directions.  
By involving more shells, one can eliminate the errors order by order.  
On the other hand, the anisotropic properties are in general very difficult 
to study.  
The isotropy of $D_4$ is more like an accident (at least in our framework) 
--- and it is straightforward to show that that the symmetry does not prevent 
higher multipoles appearing at order $a^4$.  
As a result, the most tractable way of eliminating anisotropic errors is 
by cancellation between shells related just by a simple scaling.  

We have discussed the root lattices in detail and calculated their $\Sigma$ 
explicitly.  
However, most of our results are equally valid for root and non-root lattices. 
However, root lattices are so tractable in theory, and useful in practice, 
that it makes a very natural starting point.  
The only important non-root lattice in low dimensions is $D_3^*$, the three 
dimensional body-centered cubic lattice, which is the dual of $D_3$.  
\begin{equation}
D^*_3 \equiv \{X=(x_1,x_2,x_3) \in a{\bf Z}^3 \; \cup \; 
a({\bf Z}^3 + (\textstyle{1\over2}, {1\over2}, {1\over2})) \},
\end{equation}
Its coordinates are either all integers or all half integers.  
Its nearest neighbors are $(\pm{1\over2}, \pm{1\over2}, \pm{1\over2})$, 
and angles between them are not rational fractions of $\pi$.  
One can study $D_3^*$ just as we study the root lattices, but we
do not find any properties of interest.  

What are the implications of this study for real lattice simulations?  
In no way does it challenge the canonical choice --- the hypercubical lattice, 
which always has the smallest error density.  
The only case when one may consider otherwise is when isotropy is 
important; in such scenarios the four dimensional checkerboard lattice may 
be a tempting alternative.  
Still, as disclaimed in the introduction, the geometry of the lattice is 
only one of many factors which affect the efficiency of lattice simulations.  
Algorithmic considerations may overwhelm geometrical ones; remember the two 
shell discretization of $D_4$ involves 48 neighbors, opposed to 16 for $Z_4$.  
Still, the author expects this study is useful in providing a general 
framework in which different lattices may be discussed on equal footing, 
and the results may serve as ``rules of thumb'' for lattice simulations.  

In this paper, we have studied how the geometries of non-hypercubical 
lattices affect the discretization of the Laplacian operator.  
Possible generalization to the case of the Dirac operator is possible but 
not straightforward, due to the problem of parity doubling.  
Without doing a careful analysis, it is difficult to foresee whether the 
results we have obtained in this paper can be extended to the Dirac case, 
However, the author is willing to make the following {\it conjecture}:  
the order $a^2$ discretization error of the Dirac operator does not vanish 
for any choice of lattice.  
Just as the discretization of the Laplacian operator is always plagued with 
errors like $a^2 \partial^4$ which have the same tranformation properties 
as the Laplacian $\partial^2$, it is probable that the discretization of 
the Dirac operator $\gamma \cdot \partial$ will suffer from errors like 
$a^2 (\gamma \cdot \partial) \partial^2$.  
Hopefully the author can extend the present framework to the Dirac case in 
the near future and determine the correctness of this conjecture.  

\acknowledgements
I would like to thank Silas Beane, Tom Cohen, Jim Griffin and Steve Wallace 
for discussions.  
This work is supported by the U.S.~Department of Energy grant 
DE-FG02-93ER-40762.

\end{document}